\documentstyle[aps,preprint,prb]{revtex}
\begin{document}
\tightenlines
\def\vk{\vec k} 
\def\br{{\bf r}}
\title{\bf Weak Localization and the Mooij Rule in Disorderd Metals }
\author{Mi-Ae Park }
\address{Department of Physics,  University of Puerto Rico at Humacao,\\
 Humacao, PR 00791}
\author{Kerim Savran and Yong-Jihn Kim }
\address{Department of Physics,  Bilkent University,\\
 06533 Bilkent, Ankara, Turkey}
\maketitle
\begin{abstract}

Weak localization leads to the same correction to both the conductivity
and the McMillan's electron-phonon coupling constant $\lambda$ (and $\lambda_{tr}$). 
Consequently the temperature dependence of the thermal electrical
resistivity is decreasing as the conductivity is decreasing due to
weak localization, which results in the decrease of the temperature coefficient
of resistivity (TCR) with increasing the residual resistivity.  
When $\lambda$ and $\lambda_{tr}$ are approaching zero, 
only residual resistivity part remains  
and gives rise to the negative TCR.
Accordingly, the Mooij rule is a manifestation of weak localization 
correction to the conductivity and the electron-phonon interaction. 
This study may provide a new means of probing the phonon-mechanism in exotic
superconductors.

\end{abstract}

\vspace{1pc}

\noindent

\vfill\eject
\section{\bf Introduction} 

Although weak localization has greatly deepened our understanding of the 
normal state of disordered metals,$^{1,2,3}$ its effect
on superconductivity and the electron-phonon interaction has not been
understood well.$^{2}$ Recently, it has been shown that weak localization
leads to the same correction to the conductivity and the phonon-mediated
interaction.$^{4,5}$ 
In fact, there are overwhelming numbers of experiments which support
this idea.$^{5}$ For instance, tunneling,$^{6-8}$ specific heat,$^{9}$ 
x-ray photoemission spectroscopy (XPS),$^{10}$ correlation of $T_{c}$ and 
the residual resistivity,$^{11-13}$ universal correlation of $T_{c}$ and
the resistance ratio,$^{14-16}$ and loss of the thermal electrical 
resistivity$^{17}$ with decreasing $T_{c}$ clearly
show a decrease of the electron-phonon interaction accompanying the decrease 
of $T_{c}$ with disorder. 
It is then anticipated that the electron-phonon interaction in the normal state of metals
will also be influenced strongly by weak localization. 
We expect that phonon-limited electrical resistance,  attenuation of a sound 
wave, thermal resistance, and a shift in phonon frequencies may change
due to weak localization.$^{18}$

Indeed, the Mooij rule$^{19}$ in strongly disordered metallic systems seems to 
be a manifestation of the effect of weak localization on the electron-phonon 
interaction and the conductivity. In early seventies,
Mooij found a correlation between the residual resistivity and the temperature
coefficient of resistivity (TCR). In particular, TCR is decreasing with
increasing the residual resistivity. Then it becomes negative for resistivities
above $150\mu\Omega cm$. We stress that this behavior is consistent with the 
above superconducting properties: correlation of $T_{c}$ and 
the residual resistivity,$^{11-13}$ universal correlation of $T_{c}$ and
the resistance ratio,$^{14-16}$ and loss of the thermal electrical 
resistivity$^{17}$ with decreasing $T_{c}$. 

There are already several theoretical works on this problem. 
Jonson and Girvin$^{20}$ performed numerical calculations for an Anderson model 
on a Cayley tree and found that the adiabatic phonon approximation 
breaks down in the high-resistivity regime producing the negative TCR.
Imry$^{21}$ pointed out the importance of incipient Anderson localization 
(weak localization) in the resistivities of highly disordered metals.
He argued that when the inelastic mean free path, $\ell_{ph}$, is
smaller than the coherence length, $\xi$, the conductivity increases
with temperature like $\ell_{ph}^{-1}$ and thereby leads to the negative
TCR. On the other hand, Kaveh and Mott$^{22}$ generalized the Mooij rule.
Their results are as follows: The temperature dependence of the conductivity 
of a disordered metal as a function of temperature changes slope due to 
weak localization effects, and if interaction effects are included, 
the conductivity changes its slope three times. 
G\"otze, Belitz, and Schirmacher$^{23,24}$ introduced a theory with
phonon-induced tunneling. There is also the extended Ziman theory,$^{25}$ 
and Jayannavar and Kumar$^{26}$ suggested that the Mooij rule 
can arise from strong electron-phonon interaction taking into account 
qualitatively different roles of the diagonal and off-diagonal modulations.

In this paper, we propose an explanation of the Mooij rule based
on the effect of weak localization on the electron-phonon interaction.
If we assume the decrease of the electron-phonon interaction
due to weak localization,$^{4,5}$ we can understand the decrease of TCR with 
increasing the residual resistivity.
The negative TCR is therefore due to weak localization correction to
the Boltzmann conductivity, since when TCR is approaching zero there is no 
temperature-dependent resistivity left. (This latter point is similar to  
Kaveh and Mott's interpretation.$^{22}$)  
In Sec. II, we briefly describe the Mooij rule. In Sec. III, weak localization 
correction to the McMillan's electron-phonon coupling constant $\lambda$ and 
$\lambda_{tr}$ is calculated. A possible explanation of the Mooij rule 
is given in Sec. IV, and its 
implication is briefly discussed in Sec. V. 
In particular, this study may provide a means to probe the phonon-mechanism
in exotic superconductors.

\section{The Mooij Rule}

Mooij$^{19}$ was the first to point out that the size and sign of the 
temperature coefficient of resistivity (TCR) in many disordered systems 
correlate with its residual resistivity $\rho_{0}$ as follows:
\begin{eqnarray}
d\rho/dT&>&0  \quad \rm{if} \quad \rho_{0}<\rho_{M}\nonumber\\
d\rho/dT&<&0  \quad \rm{if} \quad \rho_{0}>\rho_{M}.
\end{eqnarray}
Thus, TCR changes sign when $\rho_{0}$ reaches the Mooij resistivity $\rho_{M}\cong 150\mu\Omega cm$. 
An approximate equation for $\rho(T)$ is given by$^{2}$
\begin{equation}
\rho(T)=\rho_{0}+(\rho_{M} -\rho_{0})AT,
\end{equation}
where $A$ is a constant which depends on the material.

Figure 1 shows the temperature coefficient of resistance $\alpha$ versus 
resistivity for transition-metal alloys  obtained by Mooij. 
It is clear that $\alpha$ (and TCR) is correlated with the residual 
resistivity. 
Note that above $150\mu\Omega cm$ most $\alpha$'s are negative while
no negative $\alpha$ is found for resistivities below $100\mu\Omega cm$.
Figure 2 shows the resistivity as a function of temperature for pure Ti 
and  TiAl alloys containing 3, 6, 11, and 33\% Al. TCR is decreasing
as the residual resistivity is increasing.
For TiAl alloy with 33\% Al shows the negative TCR.
We note that the positive TCR is basically high temperature phenomenon, 
presumably related to the phonon-limited resistivity, 
whereas the negative TCR is rather low temperature behavior, probably 
connected with the residual resistivity part. 
Since this behavior is generally found in strongly disordered metals and 
alloys, amorphous metals, and metallic glasses,$^{2}$ it is called the Mooij rule.
However, the physical origin of this rule has remained unexplained until now.

\section{\bf Weak Localization Correction to The Electron-Phonon Interaction} 

Since the electron-phonon interaction in metals gives rise to both the (high 
temperature) resistivity and superconductivity, these properties are
closely related, which was noticed by many workers.$^{27-31}$
Gladstone, Jensen, and Schrieffer$^{27}$ pointed out that $\lambda$ and  
the high temperature electrical resistivity are closely related each other.
Hopfield$^{28,29}$ noted that the electronic relaxation time due to
electron-phonon interaction, as measured in optical experiments above the Debye temperature, 
should be approximately  equal to $2\pi\lambda k_{B}T/\hbar$.
He applied this idea to Nb, Mo, Al and Sn and found a good agreement with experiment.  
Grimvall$^{30}$ estimated $\lambda$ for noble metals from Ziman's high temperature 
resistivity formula. Maksimov and Motulevich$^{31}$ followed the idea of Hopfield
and estimated $\lambda$ from optical measurements for Pb, Sn, In, Al, Zn, Nb, V, 
$\rm Nb_{3}Sn$, and $\rm V_{3}Ga$, which are in good agreement with  the McMillan's coupling constant
$\lambda$ from superconductivity data.

In this Section, we show that weak localization leads to the
same correction to the conductivity, the McMillan's electron-phonon
coupling constant $\lambda$, and $\lambda_{tr}$.

\subsection{High Temperature resistivity}

At high temperatures, the phonon limited electrical resistivity is$^{32-35}$
\begin{eqnarray}
\rho_{ph}(T)&=&{4\pi mk_{B}T \over ne^{2}\hbar}\int
{\alpha_{tr}^{2}F(\omega)\over \omega}d\omega,\nonumber\\
&=& {2\pi mk_{B}T \over ne^{2}\hbar}\lambda_{tr},
\end{eqnarray}
where $\alpha_{tr}$ includes an average of a geometrical factor
$1-cos\theta_{\vk\vk'}$ and $F(\omega)$ is the phonon density of states.   
On the other hand, in the strong-coupling theory of superconductivity,$^{36,37}$
the McMillan's  electron-phonon coupling constant is defined by$^{37}$
\begin{eqnarray}
\lambda=2\int{\alpha^{2}(\omega)F(\omega)\over \omega}d\omega. 
\end{eqnarray}
Assuming $\alpha_{tr}^{2}\cong\alpha^{2}$,$^{32,38-40}$ we obtain
\begin{eqnarray}
\rho_{ph}(T)&=&{2\pi mk_{B}T \over ne^{2}\hbar}\lambda_{tr}\\
&\cong& {2\pi mk_{B}T \over ne^{2}\hbar}\lambda.
\end{eqnarray} 
Consequently the McMillan's coupling constant $\lambda$ also determines 
the size and sign of TCR.

The existence of this relationship was well confirmed theoretically and experimentally.
Table I shows the comparison of $\lambda_{tr}$ and $\lambda$ by Economou$^{38}$ for various
materials. He obtained $\lambda_{tr}$ from Eq. (5) and compared with $\lambda$,
as obtained from $T_{c}$ measurements, and/or tunneling experiments, and/or first principle  
calculations.$^{39}$
The overall agreement between $\lambda_{tr}$ and $\lambda$ is impressive. 
Grimvall estimated $\lambda$ for noble metals$^{30}$ and 
noble metal alloys$^{41}$ from Eq. (6). 
Maksimov$^{40}$ also noted the direct relation between $\lambda$ and the high temperature
resistivity.
Hayman and Carbotte$^{42}$ pointed out that information on the volume dependence of
an electron-phonon coupling strength can be obtained from high temperature resistivity.
Chakraborty, Pickett, and Allen$^{43}$ used Eq. (5) to obtain the empirical  values of $\lambda_{tr}$
for Nb, Mo, Ta, and W. They found that $\lambda_{tr}$ from 
resistance and the McMillan's coupling constant $\lambda$ from superconductivity are very similar
in magnitude for these materials.
We can also mention experimental confirmations by Rapp and Crawfoord$^{44}$ for Nb-V alloys,
Rapp and Fogelholm$^{45}$ for Al-Mg alloys,
Fl\"ukiger and Ishikawa$^{46}$ for Zr-Nb-Mo alloys, 
 Fogelholm and Rapp$^{47}$ for In-Sn alloys,
Lutz et al.$^{48}$ for $\rm Nb_{3}Ge$ films,
Man'kovskii et al.$^{49}$ for thin Sn
films, Rapp, Mota and Hoyt$^{50}$ for Au-Ga alloys, and Sundqvist and Rapp$^{51}$ 
for aluminum under pressure.
Figure 3 shows the McMillan's coupling constant $\lambda$ versus $d\rho/dT\propto \lambda_{tr}$ 
for Au-Ga, Au-Al, and Ag-Ga alloys,$^{52}$ which exemplifies the correlation implied by Eq. (6).

\subsection{Weak localization correction to the McMillan's coupling constant $\lambda$ and $\lambda_{tr}$}

Now we need to calculate the McMillan's electron-phonon coupling constant 
$\lambda$ for highly disordered systems.
We follow McMillan's approach to the strong-coupling theory.$^{37,5}$
(For simplicity we consider an Einstein model with frequency $\omega_{D}$). 
He showed that  $\lambda$ can be written as$^{37}$
\begin{eqnarray}
\lambda&=&2\int{\alpha^{2}(\omega)F(\omega)\over \omega}d\omega \\
&=&N_{0}{<I^{2}>\over M<\omega^{2}>},
\end{eqnarray}
where $M$ is the ionic mass and $N_{0}$ is the electron density of states 
at the Fermi level.  
$<I^{2}>$ is  the average over the Fermi surface of the 
square of the electronic matrix element and  
$<\omega^{2}>=\omega_{D}^{2}$.
In the presence of impurities, weak localization leads to a correction to 
$\alpha^{2}(\omega)$ or $<I^{2}>$, (disregarding the changes of $F(\omega)$ and $N_{0}$). 

There are two ways to obtain the McMillan's coupling constant $\lambda$ in the presence of impurities.
One method is to calculate $\lambda$ directly from Eq. (8), using the electronic
matrix element for disordered systems and the other is to carry out the 
canonical transformation of Fr\"ohlich in the scattered state basis.$^{5,53}$ 
We have found that both methods lead to the same $\lambda$.

In this paper, we use the latter method in a simple manner by observing that  
the Fr\"ohlich interaction can be derived from the phonon Green's function.$^{54}$ 
We note that the equivalent electron-electron potential in the 
electron-phonon problem is given by the phonon Green's function $D(x-x')$:$^{54-56}$
\begin{equation}
V(x - x') \rightarrow {I_{0}^{2}\over M\omega_{D}^{2}} D(x-x'),
\end{equation}
where  $x=({\bf r},t)$ and $I_{0}$ is the electronic matrix element for the plane wave states. 
The Fr\"ohlich interaction at finite temperatures is then given by$^{54}$
\begin{eqnarray}
V_{nn'}(\omega, \omega')&=& 
{I_{0}^{2}\over M\omega_{D}^{2}} \int\int d{\bf r}d{\bf r'}
\psi_{n'}^{*}({\bf r}) \psi_{\bar{n}'}^{*}({\bf r'})D({\bf r}-{\bf r'},\omega-\omega')
\psi_{\bar{n}}({\bf r'}) \psi_{n}({\bf r})\nonumber\\
&=& 
{I_{0}^{2}\over M\omega_{D}^{2}} \int|\psi_{n'}({\bf r})|^{2} |\psi_{n}({\bf r})|^{2}d{\bf r}{\omega_{D}^{2}\over
\omega_{D}^{2}+(\omega-\omega')^{2}}\nonumber\\
&=& V_{nn'} {\omega_{D}^{2}\over \omega_{D}^{2}+(\omega-\omega')^{2}},
\end{eqnarray}
where 
\begin{eqnarray}
D({\bf r}-{\bf r'},\omega-\omega')&=&\sum_{\vec q}{\omega_{D}^{2}\over (\omega-\omega')^{2}+\omega_{D}^{2}}
e^{i{\vec q}\cdot({\bf r}-{\bf r'})}\nonumber\\
&=& {\omega_{D}^{2}\over (\omega-\omega')^{2}+\omega_{D}^{2}}
\delta({\bf r}-{\bf r'}).
\end{eqnarray}
Here $\omega$ means the Matsubara frequency and $\psi_{n}$ and $\psi_{\bar n}$
 denote the scattered state and its time-reversed partner, respectively. 
Therefore, we get the strong-coupling gap equation$^{5}$
\begin{eqnarray}
\Delta({n}, \omega)&=&T\sum_{\omega'}\sum_{n'}
V_{{ n}{n'}}(\omega, \omega') {\Delta({n'}, \omega')\over 
\omega'^{2}+E_{n'}^{2}(\omega')}\nonumber\\
&=&T\sum_{\omega'}
{\omega_{D}^{2}\over (\omega-\omega')^{2}+\omega_{D}^{2}}
\sum_{n'}
V_{{n}{n'}} {\Delta({n'}, \omega')\over 
\omega'^{2}+E_{n'}^{2}(\omega')},
\end{eqnarray}
where
\begin{equation}
E_{n'}(\omega')=\sqrt{\epsilon_{n'}^{2}+\Delta_{n'}^{2}(\omega')},
\end{equation}
and the McMillan's electron-phonon coupling constant $\lambda$ 
\begin{equation}
\lambda=N_{0}<V_{nn'}(0,0)>=N_{0}{I_{0}^{2}\over M\omega_{D}^{2}}<\int
|\psi_{n}({\bf r})|^{2} |\psi_{n'}({\bf r})|^{2}d{\bf r}>.
\end{equation}
Here $\epsilon_{n}$ means the eigenenergy of the scattered state $\psi_{n}$.
As expected, Eq. (14) is the same as Eq. (8).
It is remarkable that the McMillan's electron-phonon coupling constant is
determined by the density correlatin function.

Note also that in the presence of impurities, the density correlation function 
has a free-particle form for $t<\tau$ (scattering time) and a diffusive
form for $t>\tau$.$^{57}$ 
As a result, for $t>\tau$ (or $r>\ell$), 
one finds$^{58-62}$ 
\begin{eqnarray}
R(t>\tau) &=& \int_{t>\tau} |\psi_{n}({\bf r})|^{2}|\psi_{n'}({\bf r})|^{2}d{\bf r} \nonumber\\ 
&=& \sum_{\vec q}|<\psi_{n}|e^{i{\vec q}\cdot {\bf r}}|\psi_{n'}>|^{2}\nonumber\\
&=&\sum_{\pi/L<\vec q<\pi/\ell}{1\over 2\pi\hbar N_{0}D{\vec q}^{2}}\\
&=&{3\over 2(k_{F}\ell)^{2}}(1-{\ell\over L}).
\end{eqnarray}
Here $\ell$ is the mean free path and $L$ is the inelastic diffusion length.
Whereas the contribution from the free-particle-like density correlation
for $t<\tau$ is$^{5,58}$  
\begin{eqnarray}
R(t<\tau) &=& \int_{t<\tau} |\psi_{n}({\bf r})|^{2}|\psi_{n'}({\bf r})|^{2}d{\bf r} \nonumber\\ 
&=&  [1-{3\over (k_{F}\ell)^{2}}(1-{\ell\over L})].
\end{eqnarray}
Since the phonon-mediated interaction is retarded for 
$t_{ret}\sim 1/\omega_{D}$, only the free-particle-like density correlation
contributes to $\lambda$. This is also true of $\lambda_{tr}$, simply because the conductivity
is determined by the behavior of the wavefunction $\psi$ for $t<\tau$ (or $r<\ell$).$^{63}$

Consequently, we obtain weak localization correction to the McMillan's 
coupling constant $\lambda$ and $\lambda_{tr}$:
\begin{equation}
\lambda=N_{0}{I_{0}^{2}\over M\omega_{D}^{2}} [1-{3\over (k_{F}\ell)^{2}}(1-{\ell\over L})].
\end{equation}
and  
\begin{eqnarray}
\lambda_{tr} &=&2\int{\alpha_{tr}^{2}(\omega)F(\omega)\over \omega}d\omega\nonumber\\ 
&\cong& {N_{0}I_{0}^{2}\over M\omega_{D}^{2}}
 [1-{3\over (k_{F}\ell)^{2}}(1-{\ell\over L})]\nonumber\\
&=& {N_{0}I_{0}^{2}\over M\omega_{D}^{2}} [1-{3\over (k_{F}\ell)^{2}}].
\end{eqnarray}
We have used the fact that $L$ is effectively infinite at $T=0$.
Note that the weak localization correction term is the same
as that of the conductivity.

\section{Explanation of the Mooij Rule}

As noted in the Section II, the positive TCR is high temperature
phenomenon whereas the negative TCR is low temperature phenomenon.
Thus, the decrease of the positive TCR is mainly due to  
the decrease of the phonon-limited resistivity, which is a manifestation
of weak localization correction to the electron-phonon interaction.  
On the other hand, the negative TCR originates from the residual 
resistivity, which is also a manifestation of weak localization
correction to the conductivity.
Accordingly, weak localization seems to be the physical origin of 
the Mooij rule in disordered metals. One should note that this observation 
agrees with the superconducting behavior of disordered systems, such as
 correlation of $T_{c}$ and the residual resistivity,$^{11-13}$ 
universal correlation of $T_{c}$ and
the resistance ratio,$^{14-16}$ and loss of the thermal resistivity$^{17}$ 
with decreasing $T_{c}$. 

\subsection{Decrease of TCR at high temperatures}

Upon substituting Eq. (19) into Eq. (3), one finds the phonon-limited high 
temperature resistivity
\begin{eqnarray}
\rho_{ph}(T)&=&{2\pi mk_{B}T \over ne^{2}\hbar}\lambda_{tr}\nonumber\\
&\cong& {2\pi mk_{B}T \over ne^{2}\hbar} {N_{0}I_{0}^{2}\over M\omega_{D}^{2}}
[1-{3\over (k_{F}\ell)^{2}}].  
\end{eqnarray} 
Note that as the disorder parameter $1/k_{F}\ell$ is increasing, 
both the magnitude of the phonon-limited resistivity and the TCR decrease.
This behavior is due to the reduction of the McMillan's electron-phonon
coupling constant when electrons are weakly localized. 
It is remarkable that the slope of the high temperature resistivity 
varies as $\sim 1/(k_{F}\ell)^{2}$, in accord with the behavior
of the residual resistivity.

The phonon-limited resistivity $\rho_{ph}$ versus temperature T is shown
in Fig. 4 (a) for six values of $k_{F}\ell$.   
We used $k_{F}=0.8\AA^{-1}$, $n=k_{F}^{3}/3\pi^{2}$,   
and $N_{0}I_{0}^{2}/(M\omega_{D}^{2})=0.5$.  
It is clear that TCR is decreasing significantly as the electrons
are weakly localized.

\subsection{Negative TCR at low temperatures}

At low temperatures the conductivity and the residual resistivity are given 
by$^{2,3}$
\begin{eqnarray}
\sigma&=&\sigma_{B}[1-{3\over (k_{F}\ell)^{2}}(1-{\ell\over L})],
\end{eqnarray}
and
\begin{equation}
\rho_{0}= {1\over  \sigma_{B}[1-{3\over (k_{F}\ell)^{2}}(1-{\ell\over L})]},
\end{equation}
where $\sigma_{B}=ne^{2}\tau/m$.
When $1/k_{F}\ell$ becomes comparable to $\sim 1$, the magnitude and slope of 
$\rho_{ph}(T)$ are negligible. In that case, only the residual 
resistivity will play an important role. 
Therefore, the  observed  negative TCR may be understood from the residual 
part. With decreasing $T$, 
since the inelastic diffusion length $L$ increases,
the residual resistivity will also increase, leading to the negative TCR.
We stress that both the phonon-limited resistivity and the residual 
resistivity have the same quadratic dependence on the disorder parameter 
$1/k_{F}\ell$.

Figure 4 (b) shows the temperature dependence of the residual resistivity  
$\rho_{0}$ for $k_{F}\ell=2.2, 2.4, 2.8, 3.4, 5,$ and $15$. 
Since it is difficult to evaluate $k_{F}\ell$ up to a factor of 2,$^{64}$
we assumed that $\rho_{0}=100\mu\Omega cm$ corresponds to $k_{F}\ell=3.2$.
We used the same $k_{F}$ as in Fig. 4 (a) and $L=\sqrt{D\tau_{i}}=\sqrt{\ell}\times 350/T (\AA)$.
Here $D$ is the diffusion constant and $\tau_{i}$ denotes the inelastic 
scattering time.
When $k_{F}\ell$ is comparable to 1,  the negative TCR emerges.
Notice the scale difference between Figures, 4 (a) and 4 (b).

\subsection{Comparison with experiment}

In Sections IV. A and B, we have explained the physical
origin of the Mooij rule. In this section, we compare our theoretical
resistivity curve and the experimental data (Figure 2) for extended
temperature range.
Let us remind the approximate formula for $\rho(T)$ suggested by Lee and 
Ramakrishnan, i.e.,$^{2}$
\begin{equation}
\rho(T)=\rho_{0}+(\rho_{M} -\rho_{0})AT.
\end{equation}
This form of equation can be obtained if we add the residual resistivity Eq. (22)
and the phonon-limited resistivity Eq. (20), that is,
\begin{eqnarray}
\rho(T)&=&\rho_{0}+\rho_{ph}(T)\nonumber\\
&=&
 {1\over  \sigma_{B}[1-{3\over (k_{F}\ell)^{2}}(1-{\ell\over L})]}
+{2\pi mk_{B}T \over ne^{2}\hbar} 
{N_{0}I_{0}^{2}\over M\omega_{D}^{2}} [1-{3\over (k_{F}\ell)^{2}}].  
\end{eqnarray} 
It should be noticed that the addition of both resistivities does not mean the 
Matthiessen's rule.
Here we included the interference effect between the electron-phonon and electron-impurity
interactions: 
\begin{equation}
\rho(T)=\rho_{0}+\rho_{ph}(T, c=0)+\Delta \rho_{ph}^{int},
\end{equation}
where $c$ denotes an impurity concentration.
Whereas Altshuler$^{65}$ and Reizer and Sergeev$^{66}$ investigated corrections to the
impurity resistivity due to the interference, we have considered its correction
to the phonon-limited resistivity.
Since the interference correction to the impurity resistivity is $\sim 1\%$ of the
residual resistivity,$^{66, 67}$ we neglect its effect for simplicity.

In general, the phonon-limited resistivity at any temperature $T$ is given by$^{32-35}$
\begin{equation}
\rho_{ph}(T)={4\pi m \over ne^{2}}\int
{(\beta\hbar \omega)\alpha_{tr}^{2}(\omega)F(\omega)\over (e^{\beta\hbar\omega}-1)(1-e^{-\beta\hbar\omega})}d\omega,
\end{equation}
where $\beta=1/k_{B}T$. For an Einstein phonon model with$^{68}$ 
\begin{equation}
\alpha_{tr}^{2}(\omega)F(\omega)= {N_{0}I_{0}^{2}\over 2M\omega_{D}}\delta(\omega-\omega_{D}),   
\end{equation}
it is rewritten as$^{69}$
\begin{equation}
\rho_{ph}(T)={2\pi m \over ne^{2}}
 {N_{0}I_{0}^{2}\over M\omega_{D}^{2}}   
{(\beta\hbar \omega_{D})\omega_{D}\over (e^{\beta\hbar\omega_{D}}-1)(1-e^{-\beta\hbar\omega_{D}})}.
\end{equation}
It is necessary to emphasize that this result is exact for the phonon-limited resistivity in an Einstein model.$^{67}$
Including the weak localization correction to $\alpha^{2}(\omega)\cong\alpha_{tr}^{2}(\omega)$,
\begin{equation}
\alpha_{tr}^{2}(\omega)F(\omega)= {N_{0}I_{0}^{2}\over 2M\omega_{D}} [1-{3\over (k_{F}\ell)^{2}}]  
\delta(\omega-\omega_{D}),   
\end{equation}
one finds
\begin{equation}
\rho_{ph}(T)={2\pi m \over ne^{2}}
 {N_{0}I_{0}^{2}\over M\omega_{D}^{2}} [1-{3\over (k_{F}\ell)^{2}}]  
{(\beta\hbar \omega_{D})\omega_{D}\over (e^{\beta\hbar\omega_{D}}-1)(1-e^{-\beta\hbar\omega_{D}})}.
\end{equation}
Finally, we obtain the total resistivity at any temperature $T$:
\begin{eqnarray}
\rho(T)&=&\rho_{0}+\rho_{ph}(T)\nonumber\\
&=&
 {1\over  \sigma_{B}[1-{3\over (k_{F}\ell)^{2}}(1-{\ell\over L})]}
+
{2\pi m \over ne^{2}}
 {N_{0}I_{0}^{2}\over M\omega_{D}^{2}} [1-{3\over (k_{F}\ell)^{2}}]  
{(\beta\hbar \omega_{D})\omega_{D}\over (e^{\beta\hbar\omega_{D}}-1)(1-e^{-\beta\hbar\omega_{D}})}.
\end{eqnarray} 
(If we consider the Debye and realistic phonon models, there are minor changes. However, the overall
behavior is the same. More details will be published elsewhere.)

Figure 5 shows the resistivity as a function of temperature
 for $k_{F}\ell=2.3, 2.5, 2.8, 3.4, 5,$ and $15$. 
The solid lines represent the resistivity from an accurate expression Eq. (31), 
while the dashed lines are obtained from Eq. (24). 
We used the same parameters as those in Fig. 4 and $\hbar\omega_{D}=250K$.   
It is noteworthy that both equations give rise to almost the same curve as the
system is more disordered. 
For low temperatures $\tau_{i}$ is determined by electron-electron scattering
while for high temperatures it is determined by the electron-phonon scattering. 
Since we are interested in rather high temperatures, 
we assumed $\tau_{i}\sim T^{-1}$ corresponding to the electron-phonon 
scattering.$^{2,3}$  
Considering the crudeness of our calculation, the overall behavior is in good 
agreement with experiment, Fig. 2.

\section{\bf Discussion} 

At low temperatures the interference of the Coulomb interaction and the
impurity scattering leads to the interaction correction to the 
conductivity.$^{60,2}$ This effect is described by$^{70}$
\begin{equation}
\sigma= \sigma_{B}[1-{3\over (k_{F}\ell)^{2}}(1-{\ell\over L})
-{C\over (k_{F}\ell)^{2}}(1-{\ell\over L_{T}})],
\end{equation}
where $L_{T}=(\hbar D/k_{B}T)^{1/2}$ and $C\sim 1.$ 
The second correction term is the interaction term.
The constant $C$, however, changes sign depending on the exchange and 
Hartree terms and since it is difficult to determine $C$,$^{2,3,70}$ 
we did not include this term.
But it may be important at much lower temperatures. 

It is clear that weak localization effect on the electron-phonon interaction
needs more theoretical and experimental studies. In particular, weak 
localization effect on the attenuation of a sound wave, shear modulus, 
thermal resistance, and a shift in phonon frequencies will be very interesting.
Since superconductivity is also caused by the electron-phonon interaction,
comparative study of  the normal and superconducting properties of the metallic
samples will be beneficial. There is already compelling evidence that this
is the case. For instance, Testardi and his coworkers$^{14-17}$ found the 
universal correlation of $T_{c}$ and the resistance ratio. They also found that
decreasing $T_{c}$ is accompanied by the decrease of the thermal electrical
resistivity.$^{17}$ 

Observe that this study may provide a means of probing the phonon-mechanism
in exotic superconductors, such as, heavy fermion superconductors,
organic superconductors, fullerene superconductors,  and high $T_{c}$ cuprates. 
For superconductors caused by the electron-phonon interaction we expect the
following behavior. As the electrons are weakly localized by impurities or radiation damage, 
the electron-phonon interaction is weakened. As a result, both $T_{C}$
and TCR are decreasing at the same rate. When $\lambda$ is approaching zero,
both $T_{c}$ and TCR drops to zero almost simultaneously. 
When this happens we may say that 
the electron-phonon interaction is the origin of the pairing in the 
superconductors.
This behavior was already confirmed in A15 superconductors$^{14-17}$ and 
Ternary superconductors.$^{71}$
More details will be published elsewhere.

\section{\bf Conclusion} 

It is shown that weak localization decreases 
both the conductivity and the electron-phonon interaction at the same rate 
and thereby leads to  the Mooij rule. 
As the residual resistivity is increasing due to weak localization, so
the thermal electrical resistivity is decreasing,
producing the decrease of TCR. When the electron-phonon interaction is
near zero, only the residual resistivity is left and therefore
the negative TCR obtains. 
This study may provide a means of probing the phonon-mechanism in exotic
superconductors, such as, heavy fermion superconductors, organic 
and fullerene superconductors, and high $T_{c}$ superconductors.

\vspace{1pc}

\centerline{\bf ACKNOWLEDGMENTS}

Y.J.K. is grateful to Prof. Bilal Tanatar for discussions and
encouragement. M.P. thanks the FOPI at the University of Puerto Rico-Humacao
for release time.

\vfill\eject

{\bf Table I.} Comparison of $\lambda_{tr}$ and  the McMillan's electron-phonon coupling 
constant $\lambda$. Data are from Economou, Ref. 38 and Grimvall, Ref. 39.

\vspace{2pc}
\hspace{2pc}

\begin{tabular}{lrrlrr} \hline \hline
{Metal}\hspace{2pc}  & $\lambda_{tr}$\hspace{2pc} & \hspace{2pc} {$\lambda$}& 
\hspace{4pc}{Metal}\hspace{2pc}  & $\lambda_{tr}$\hspace{2pc} & \hspace{2pc} {$\lambda$}  \\ \hline

Li \hspace{2pc} & .40 \hspace{2pc} & \hspace{1pc} .41$\pm$.15 & \hspace{4pc} Na  \hspace{2pc} & .16 \hspace{2pc} & \hspace{1pc} .16$\pm$.04\\ 

K \hspace{2pc}  & .14 \hspace{2pc} & \hspace{1pc} .13$\pm$.03 & \hspace{4pc} Rb \hspace{2pc} & .19 \hspace{2pc} & \hspace{1pc} .16$\pm$.04\\ 

Cs \hspace{2pc} & .26 \hspace{2pc} & \hspace{1pc} .16$\pm$.06 & \hspace{4pc} Mg \hspace{2pc} & .32 \hspace{2pc} & \hspace{1pc} .35$\pm$.04\\ 

Zn \hspace{2pc} & .67 \hspace{2pc} & \hspace{1pc} .42$\pm$.05 & \hspace{4pc} Cd \hspace{2pc} & .51 \hspace{2pc} & \hspace{1pc} .40$\pm$.05\\ 

Al \hspace{2pc} & .41 \hspace{2pc} & \hspace{1pc} .43$\pm$.05 & \hspace{4pc} Pb \hspace{2pc} & 1.79 \hspace{2pc} & \hspace{1pc} 1.55\\ 

In \hspace{2pc} & .85 \hspace{2pc} & \hspace{1pc} .805 & \hspace{4pc} Hg \hspace{2pc} & 2.3 \hspace{2pc} & \hspace{1pc} 1.6\\ 

Cu \hspace{2pc} & .13 \hspace{2pc} & \hspace{1pc} .14$\pm$.03 & \hspace{4pc} Ag \hspace{2pc} & .13 \hspace{2pc} & \hspace{1pc} .10$\pm$.04\\ 

Au \hspace{2pc} & .08 \hspace{2pc} & \hspace{1pc} .14$\pm$.05 & \hspace{4pc} Nb \hspace{2pc} & 1.11 \hspace{2pc} & \hspace{1pc} .9$\pm$.2\\ \hline \hline

\end{tabular}

\vfill\eject

\begin{figure}
\caption{ The temperature coefficient of resistance $\alpha$ versus resistivity for bulk alloys (+), thin films ($\bullet$), and amorphous (X) alloys. Data are from Mooij, Ref. 19. }
\end{figure}

\begin{figure}
\caption{ Resistivity versus temperature for Ti and TiAl alloys containing 
0, 3, 6, 11, and 33\% Al. Data are from Mooij, Ref. 19. }
\end{figure}

\begin{figure}
\caption{ McMillan's coupling constant $\lambda$ versus $d\rho/dT\propto \lambda_{tr}$ for Ag-Ga, Au-Al, and Au-Ga alloys.  
Data are from Rapp, Ref. 52 and Grimvall, Ref. 32. }
\end{figure}

\begin{figure}
\caption{ (a) Phonon-limited resistivity $\rho_{ph}$ versus T for $k_{F}\ell=$ 15, 5, 3.4, 2.8, 2.4, and 2.2. (b) residual resistivity $\rho_{0}$ versus T for the same six values of $k_{F}\ell$. \hspace{2pc} }
\end{figure}

\begin{figure}
\caption{  Calculated resistivity versus temperature for $k_{F}\ell=$ 15, 5, 
3.4, 2.8, 2.5, and 2.3. The solid lines are $\rho(T)$ from an accurate formula, Eq. (31). The dashed lines represent the resistivity obtained from the approximate expression, Eq. (24).   \hspace{2pc} }
\end{figure}

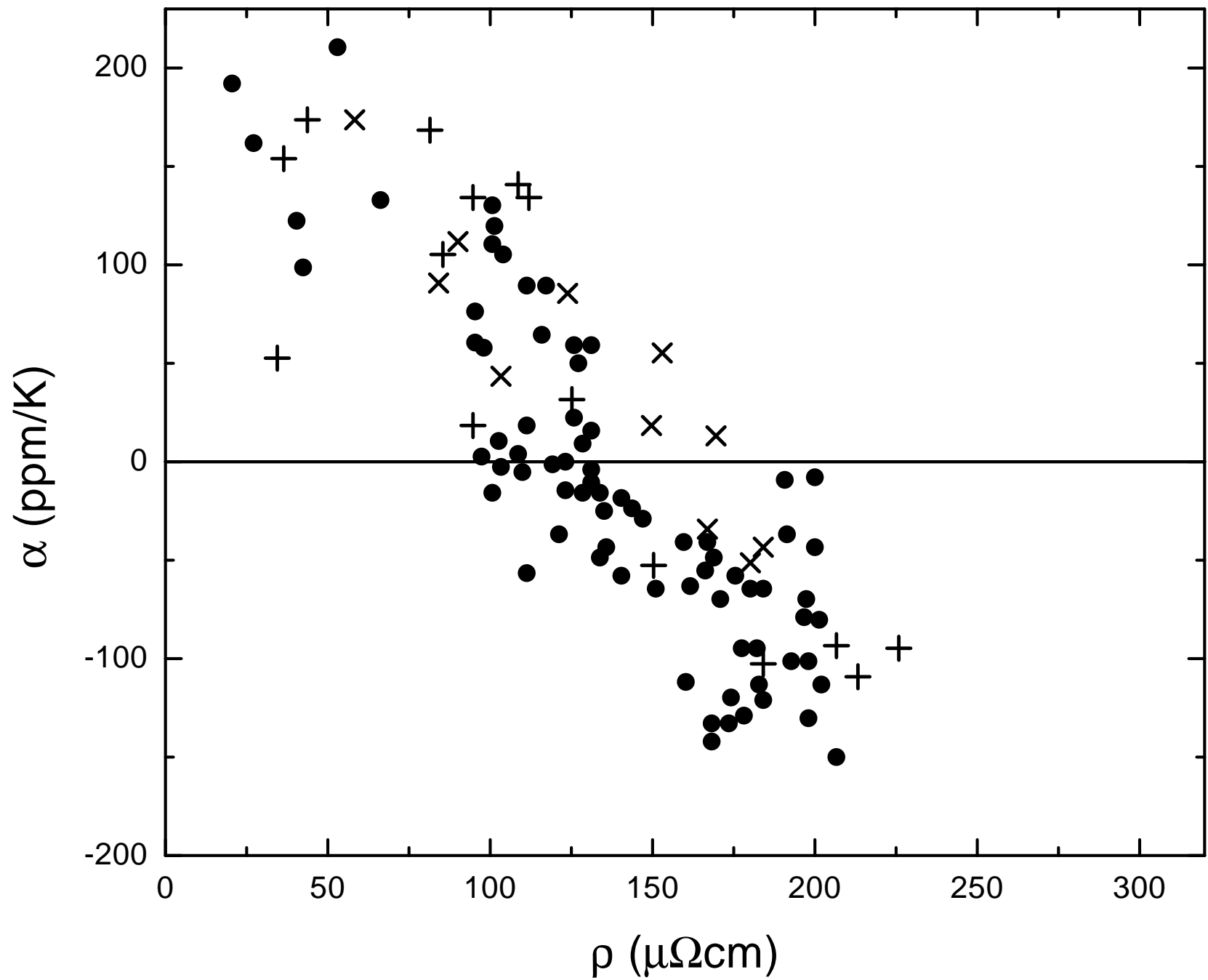
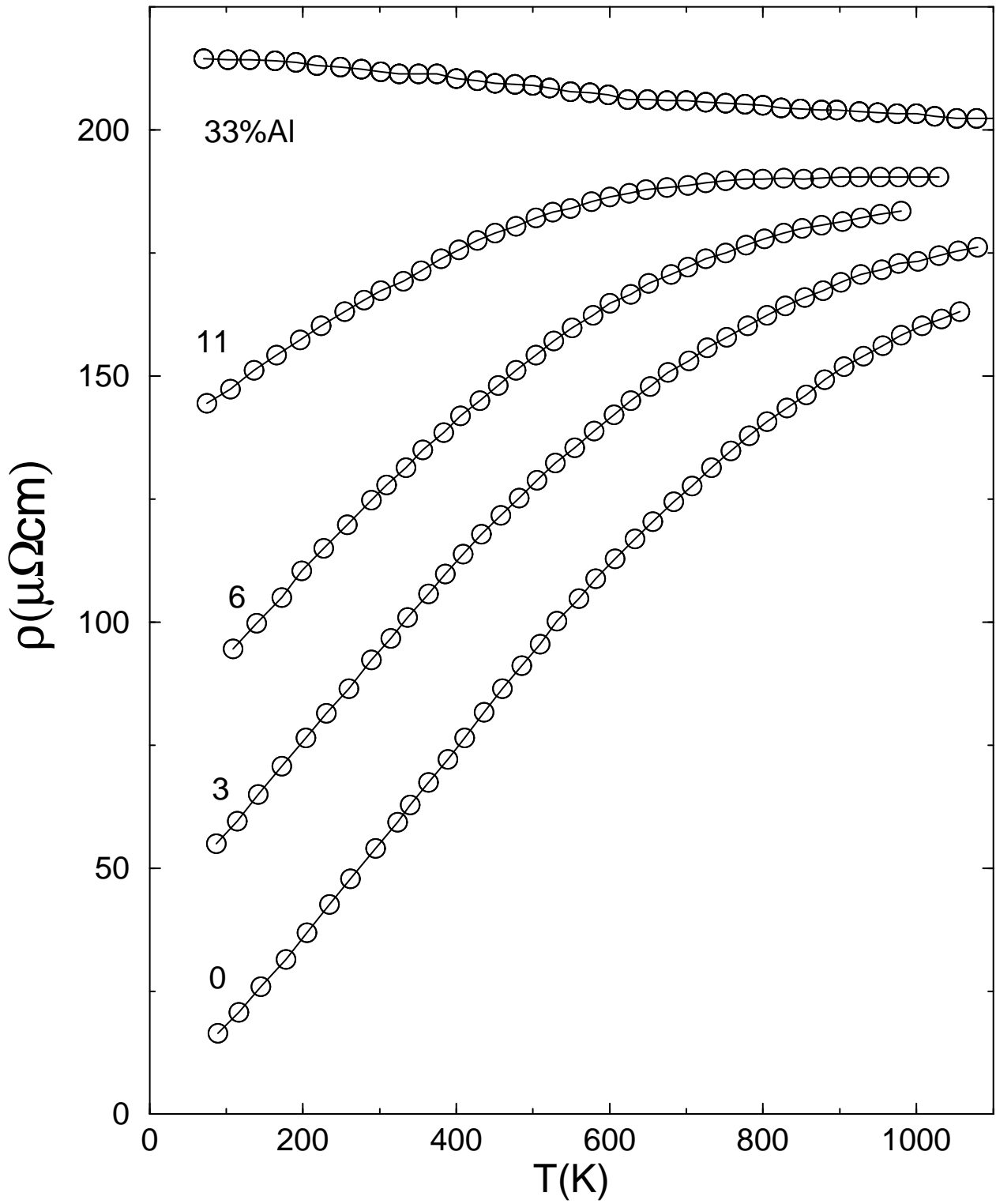
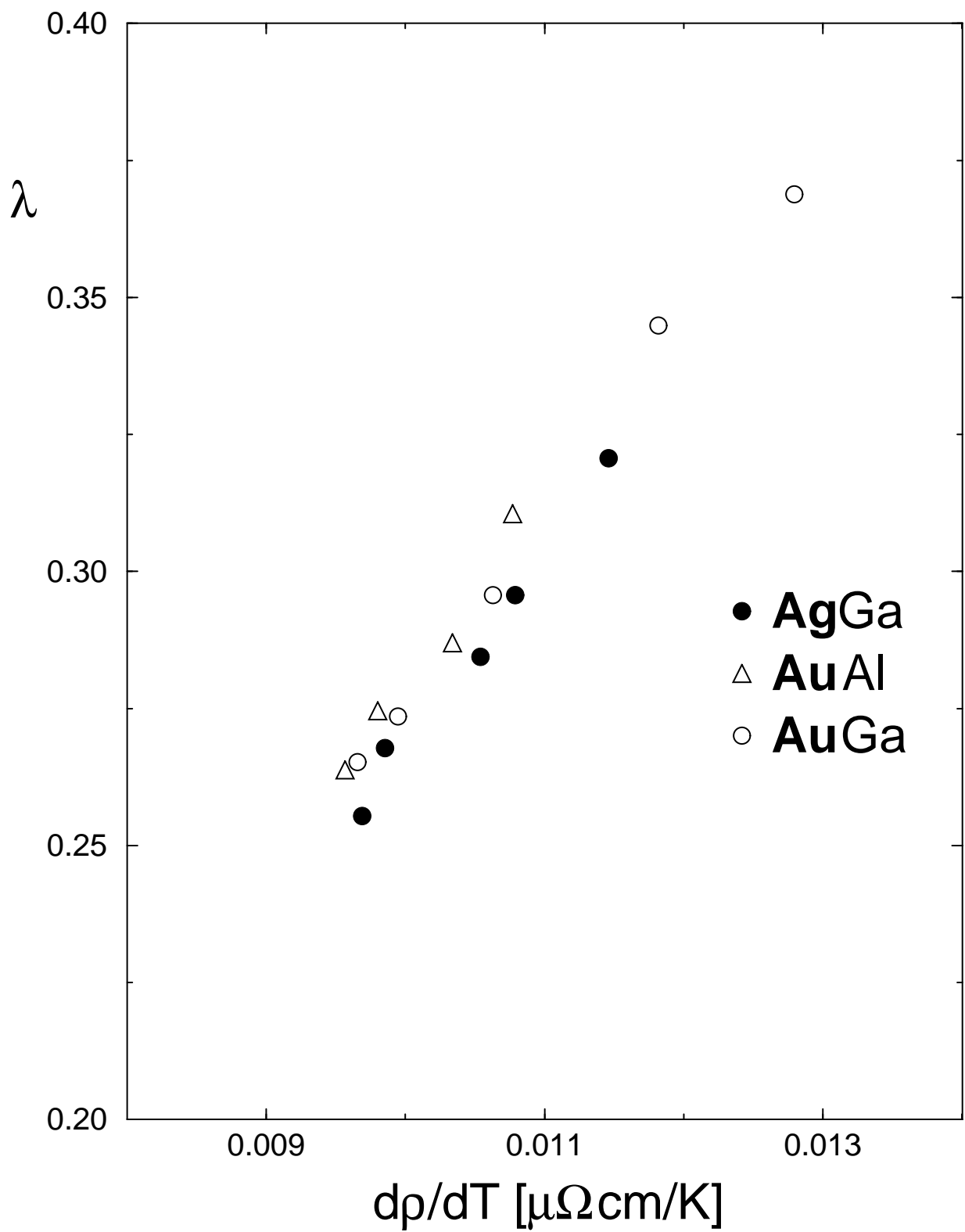
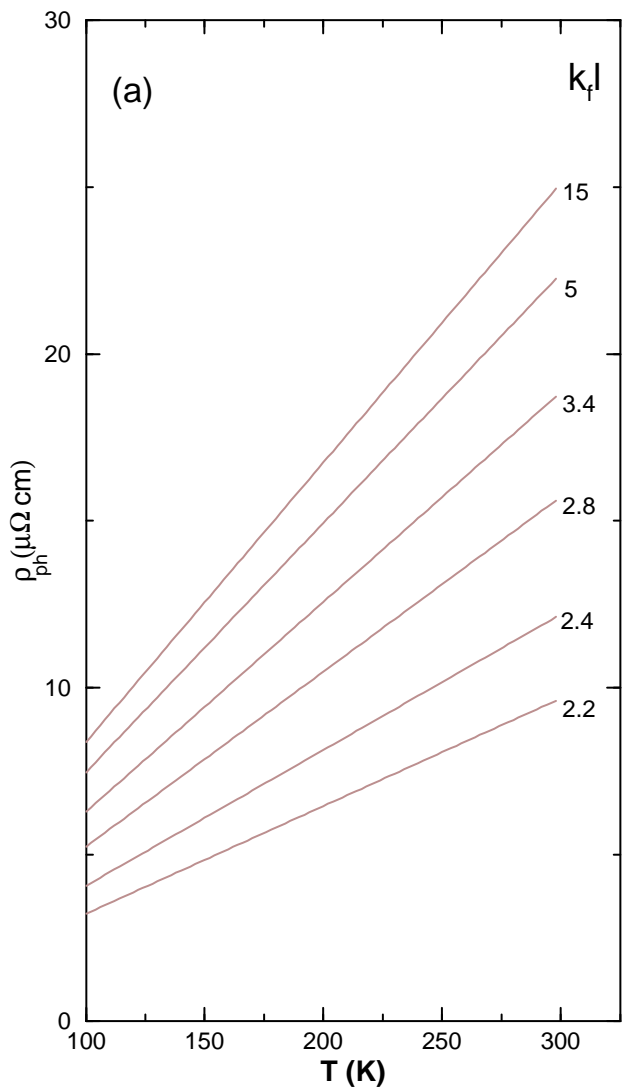
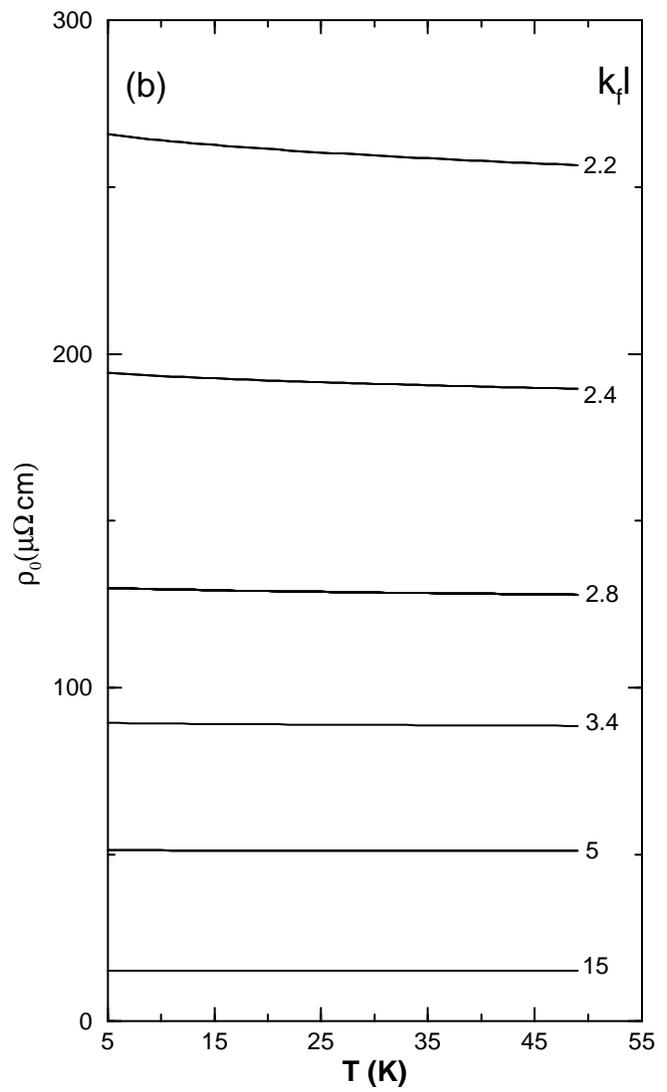
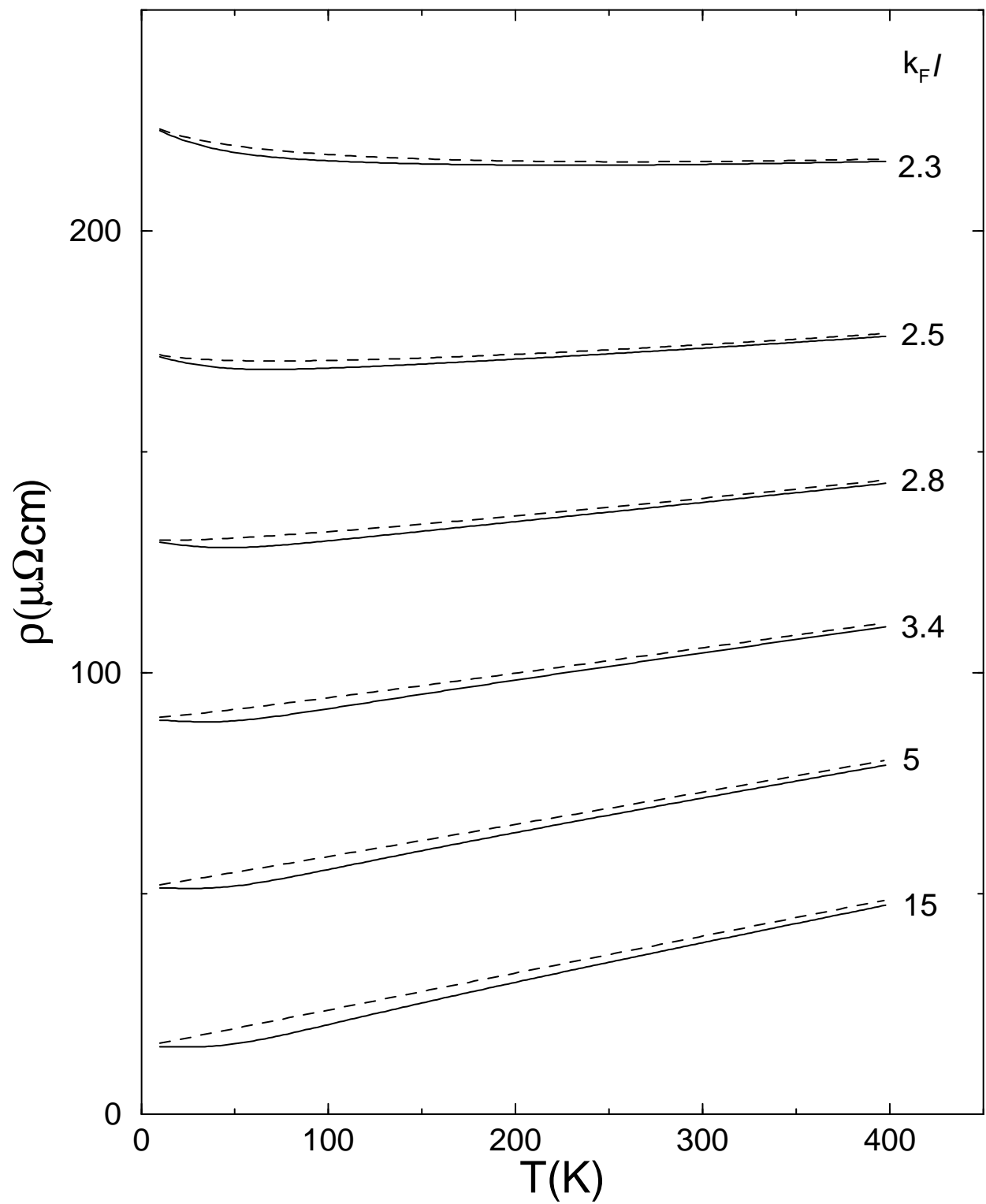
\end{document}